\documentclass{aa}

\newcommand{\grs}    {GRS 1758$-$258}
\newcommand{\unoe}   {1E 1740.7$-$2942}

\newcommand{\ltsima} {$\; \buildrel < \over \sim \;$}
\newcommand{\simlt}  {\lower.5ex\hbox{\ltsima}}            
\newcommand{\gtsima} {$\; \buildrel > \over \sim \;$}
\newcommand{\simgt}  {\lower.5ex\hbox{\gtsima}}            

\usepackage{graphicx}
\usepackage{txfonts}
\begin{document}

   \title{Optical spectroscopy of the microquasar \grs: \\
   a possible intermediate mass system?}


   \author{Josep Mart\'{\i}\inst{1,3}
          \and
          Pedro L. Luque-Escamilla\inst{2,3}  
          \and
          \'Alvaro J. Mu\~noz-Arjonilla\inst{3}
          }

\institute{
Departamento de F\'{\i}sica, Escuela Polit\'ecnica Superior de Ja\'en, Universidad de Ja\'en, Campus Las Lagunillas s/n, A3, 23071 Ja\'en, Spain\\
   \email{jmarti@ujaen.es}
   \and
Departamento de Ingenier\'{\i}a Mec\'anica y Minera, Escuela Polit\'ecnica Superior de Ja\'en, Universidad de Ja\'en, Campus Las Lagunillas s/n, A3, 23071 Ja\'en, Spain\\
  \email{peter@ujaen.es}   
  \and
  Grupo de Investigaci\'on FQM-322, Universidad de Ja\'en, Campus Las Lagunillas s/n, A3, 23071 Ja\'en, Spain\\
  \email{ajmunoz@ujaen.es}
  }

  \date{Received September 1st, 2016; accepted  October 31th, 2016}
    
 
  \abstract
   {\grs\ is one of two prototypical microquasars towards the Galactic Center direction discovered almost a quarter of a century ago. The system
   remains poorly studied in the optical domain due to its counterpart being a very faint and absorbed target in a crowded region of the sky.
   }
   {Our aim is to investigate \grs\ in order to shed light on the nature of the stellar binary components. In particular, the main physical
   parameters of the donor star, such as the mass or the spectral type, are not yet well constrained.}
   { \grs\ has remained so far elusive to optical spectroscopy owing to its observational difficulties. Here,
   we use  this traditional tool of stellar astronomy at
   low spectral resolution with a 10 m class telescope and a long slit spectrograph.
   }
   {An improved spectrum is obtained as compared to previous work. The quality of the data does not allow  the detection of emission or absorption
   features but, nevertheless, we manage to partially achieve our aims  comparing the de-reddened continuum with the spectral energy distribution expected from 
   an irradiated disc model and  different donor star templates.}
   {We tentatively propose that \grs\ does not host a giant star companion. Instead, a main sequence star with mid-A spectral type
   appears to better agree with our data. The main impacts of this finding are the possibility that we are dealing with an intermediate mass system and, in this case,
   the prediction of an orbital period significantly shorter than previously proposed.}

\keywords{  Stars: individual: \grs --- X-rays: binaries --- infrared: stars}


   \maketitle
%

\section{Introduction}

\grs, and its twin, \unoe, are  two microquasars with bipolar  jets dominating the hard X-ray appearance of the sky in the
vicinity of the Galactic Center \citep{1992ApJ...401L..15R, 1992Natur.358..215M}. Their original discovery was one of the main results
from the coded mask SIGMA telescope on board the Russian satellite GRANAT \citep{1991A&A...247L..29S, 1994Natur.371..589G}.
For many years, controversy over their actual membership to the Milky Way could not be settled until clear evidence of morphological
variability in their arc-minute extended jets was observed with modern radio interferometers in both systems \citep{2015A&A...578L..11M, 2015A&A...584A.122L}.
In the \grs\ case, the distance upper limit based on causality arguments is estimated to be $\sim 12$ kpc. 
For discussion purposes, hereafter we assume a distance of 8.5 kpc  similar to that of the Galactic Center.

Since  \grs\ is a very bright source in X-rays (see e.g. \citealt{2011MNRAS.415..410S}), it has been easily studied in this energy range.
However, this source has remained elusive to optical and near infrared analysis   owing to the significant interstellar absorption towards it.
The quest for a counterpart at these wavelengths has involved the hard work of many authors, starting with
\citet{1991Msngr..66...10B} and continued by, among others,   \citet{1998A&A...338L..95M}, \citet{ 2002ApJ...580L..61R}, and \citet{2010A&A...519A..15M}.
Only in recent times has a reliable counterpart candidate been finally identified based on both astrometric and photometric variability criteria \citep{2014ApJ...797L...1L}.
From X-ray spectral fitting, \cite{2011MNRAS.415..410S} derive an equivalent hydrogen column density
of $N_H \simeq 1.6 \times 10^{22}$ cm$^{-2}$. The corresponding  extinction at visual wavelengths can be approximately
estimated as $A_V \simeq 5.3 \times 10^{-22} {\rm cm}^{-2} N_H$ \citep{1978ApJ...224..132B}. This yields $A_V = 8.9$ mag, and this is the value
that we adopt here for  de-reddening of the spectral energy distribution (SED).

Confirmation of the stellar nature of \grs\ using spectroscopic tools with ground-based telescopes has remained a difficult task.
This is because we are dealing with a very weak counterpart
($R = 22.6\pm 0.3$, $I=21.1 \pm 0.3$,  \citealt{1998A&A...338L..95M}) in a crowded field, thus requiring the use of
 sub-arcsecond angular resolution as in \citet{2014ApJ...797L...1L}.
Previous attempts by our team to secure an optical spectrum of \grs\  under average seeing conditions using a 10m-class telescope
resulted in a poor quality result \citep{2010A&A...519A..15M}. In this paper, we report the outcome of our second attempt to acquire optical
spectroscopy of this microquasar under better observing conditions. Although the results are not yet optimal, new information is obtained that allows
us to shed more light on our physical understanding of this system.

\section{Observations}

\grs\ was observed with the {\it Gran Telescopio Canarias} (GTC) and its 
OSIRIS imager and spectrograph. The GTC run  took place on the night of 1st June 2016 from the Observatorio del Roque de los Muchachos
in the island of La Palma (Spain), under proposal code GTC23-15A. Observing conditions were clear and dark.
The R300R grism was selected to acquire a low resolution ($\lambda / \Delta \lambda \simeq 350$) spectrum.
This choice was mandatory since we are dealing with a very faint and absorbed
target whose brightness increases towards redder wavelengths.
To avoid cosmic ray excesses, the total integration time was split into different observing blocks (OBs).
Four of them could be executed with 1910 s of exposure each.
The total on-source time thus amounted  to approximately 2.1 hour in service mode. The four frames acquisition took place during the UT  time interval
01:42 to 03:54 centered around the target meridian transit. From La Palma, the source air mass does not drop below approximately 1.7,
even under these optimal conditions. Therefore, this GTC observation was a difficult one and, consequently, we requested very
good seeing conditions for the execution of our OBs. At the time of observation, the seeing at the zenith was measured
to be approximately 0.6 arc-second in the visual domain while, at the \grs\ air mass, this translated to a poorer 0.9 arc-second value. 
Assuming  a  $\propto \lambda^{-1/5}$ scaling  \citep{1978JOSA...68..877B},
this corresponds to a seeing value approaching 0.8 arc-second at our central wavelength ($0.8$ $\mu$m aprox.).
This  was judged as the minimum required to avoid
most of the contamination from a nearby star located  1.5 arc-second away from our target by using a conservative slit width of 0.6 arc-second.
The slit was also oriented along a position angle of $-16^{\circ}$, that is, orthogonal to the direction of the contaminating star
brighter than our target by  $\sim1.6$ mag in  the $R$-band. This object is also
the same K giant star previously considered a \grs\ counterpart  candidate,  but later ruled out by \citet{2014ApJ...797L...1L}.

The raw frames were processed using the IRAF software package,
 including bias and spectroscopic flat field corrections.
Wavelength calibration was secured by acquiring several arcs using different lamps (HgAr, Ne and Xe) interleaved between the individual science frames.
Target spectra were carefully extracted with sky background removed and later median combined to better exclude cosmic ray impacts.
An approximate flux calibration was applied using one of the preferred GTC spectrophotometric standards (Ross640). 
Its uncertainty is conservatively estimated to be at the $\pm 0.1$ mag level.
The slit transmission factor for the target narrow slit can be computed as:
\begin{equation}
T_x = erf  \left[  \frac{0.5 {\rm slit}}{ \sqrt{2} \frac{{\rm seeing}}{2.35482}}  \right]=0.61, 
\end{equation}
used to account for slit-losses in the target spectrum.
 
\begin{figure}
   \centering
   \includegraphics[angle=0,width=10.0cm]{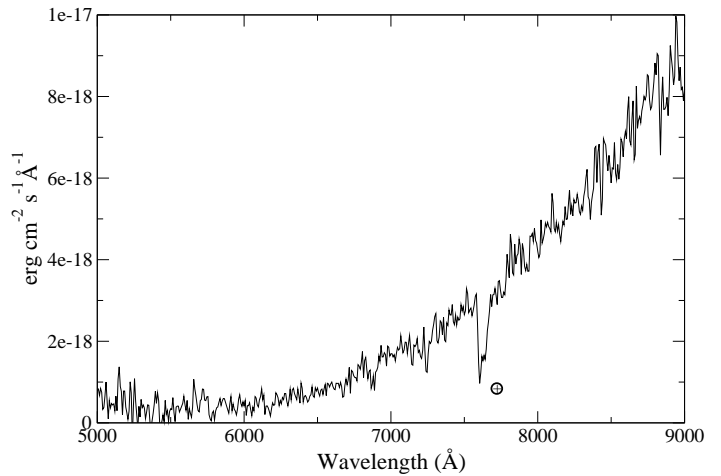}
      \caption{GTC+OSIRIS spectrum of \grs\ obtained with the R300R grism. The microquasar is barely detected at wavelengths
      shorter that 6000 \AA. The prominent absorption feature at approximately 7600 \AA\  marked with the Earth symbol $\oplus$ is of telluric origin.
    }
         \label{GTCsp}
   \end{figure}

 The final result is presented in Fig. \ref{GTCsp} where a a highly absorbed featureless continuum is the most outstanding characteristic.
To verify  the flux calibration, at the 0.8 $\mu$m wavelength the measured flux density is $4.5 \times 10^{-18}$  erg  s$^{-1}$ cm$^{-2}$  \AA$^{-1}$
 with a signal-to-noise ratio of approximately 20,
 corresponding to an $I$-band magnitude of  $21.0\pm 0.1$. 
 This corresponds closely to previous  broad-band photometry \citep{1998A&A...338L..95M}.
 
   \section{Discussion}
   
   Inspection of Fig. \ref{GTCsp} shows that no intrinsic emission nor absorption feature is reliably detected on the GTC continuum, even in the
   H$\alpha$ region where these features were more highly expected.  The quality of the spectrum is thus not adequate to reveal stellar
   features that could enable a direct determination of the companion's spectral type. Nevertheless, the overall properties of the
   continuum level do provide some useful information when considered in the context of the system broad-band SED.
   For this purpose,   the observed spectrum in Fig. \ref{GTCsp} has been de-reddened using the $A_V$ value quoted in the introduction.
   When doing so, we have kept only the data points where the microquasar counterpart was detected with a signal-to-noise ratio
   higher than 5 ($\lambda \gtrsim 7000$ \AA).

   \begin{figure}
   \centering
   \includegraphics[angle=0,width=10.0cm]{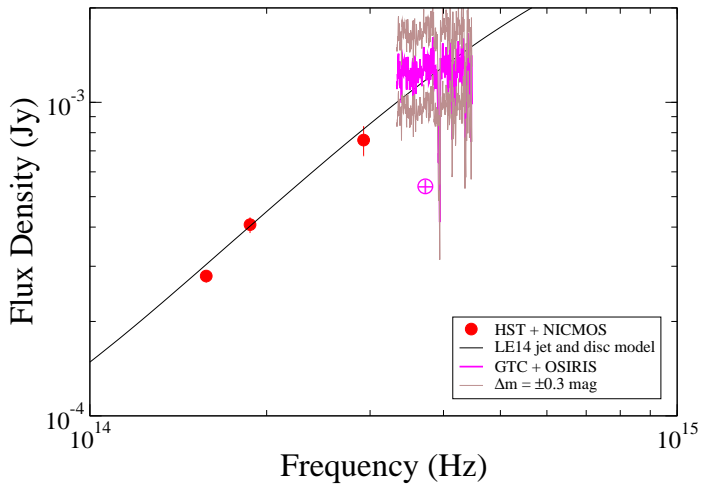}
      \caption{De-reddened GTC spectrum of \grs\ plotted together with  HST-NICMOS observations and the irradiated disc model by \citet{2014ApJ...797L...1L}
      labelled as LE14. For comparison purposes,
      the GTC spectrum has also been plotted in brown scaled by a factor corresponding to variability
      and absolute flux uncertainty effects with $\Delta m=\pm 0.3$ mag.
      The symbol $\oplus$ denotes the strongest telluric absorption feature.  
          }
         \label{sed1}
   \end{figure} 

For comparison purposes, Fig. \ref{sed1} shows the de-reddened GTC spectrum together with the irradiated disc model and the {\it Hubble} Space Telescope (HST) observations as reported in 
    \citet{2014ApJ...797L...1L}. 
    Their {\it HST}-NICMOS flux densities 
   indicated a near-infrared spectrum rising with frequency
   according to $S_{\nu} = (0.13 \pm 0.01 {\rm mJy}) \left[\nu / 10^{14}~{\rm Hz}   \right]^{+1.7 \pm 0.2}$.
   This was interpreted as mostly due to emission
   from an irradiated accretion disc. In this context,  the non-degenerated stellar companion was assumed to provide an
   almost negligible flux contribution, as is usual for low-mass X-ray binaries. However, extrapolation of this power law at higher optical frequencies does not appear now
   to be  consistent with the much flatter GTC-OSIRIS spectrum (see  again Fig. \ref{sed1}). 
   The non-detection of a steeply rising spectrum in the GTC window forces
   us to revise our previous interpretation and points to the stellar continuum actually playing a role in the near infrared and optical domain.

The apparent mismatch between the {\it HST}-NICMOS points and the GTC spectrum in Fig. \ref{sed1} 
could be due to the combined effect of the $\sim 0.2$ mag observed photometric variability \citep{2014ApJ...797L...1L} and the uncertainty
in the flux calibration ($\pm 0.1$ mag). Therefore,  in this figure we also display the GTC data with scaling factors equivalent to $\Delta m = \pm 0.3$ mag.
The lower brightness level would provide a much better continuity between the non simultaneous infrared and optical data.
The uncertainty associated with the corresponding scaling factor ($10^{-0.4\Delta m}$) is approximately 5\% for a GTC spectrum with signal-to-noise ratio of $\sim20$.
Hereafter, we assume that this down-scaled spectrum is the most plausible data set for a physical discussion.

   We carried out a tentative exploration of  possible stellar companions 
   using the Castelli and Kurucz atlas\footnote{{\tt http://www.stsci.edu/hst/observatory/crds/\\castelli\_kurucz\_atlas.html}}.
    The same jet and irradiated disc model parameters used by \citet{2014ApJ...797L...1L} were initially adopted, and part of the calculations were
   carried out using the Xspec software package from NASA.
   In order to achieve an acceptable fit, we tried different main sequence spectral types from B to M and different outer disc radii from 
   $10^{3.5}$  to $10^3$ times the inner radius. Reducing the outer radius, always within the Xspec-allowed limits, 
   removes part of the disc cooler regions that contribute significantly to optical and infrared emission. 
   This enables a more dominating role of the stellar continuum in the 0.5-2 $\mu$m region.
   The remaining parameters, including the 8.5 kpc distance, were kept identical to those of \citet{2014ApJ...797L...1L} since 
   extensive modelling is beyond the approach of this observational paper.

     \begin{figure}
   \centering
   \includegraphics[angle=0,width=10.0cm]{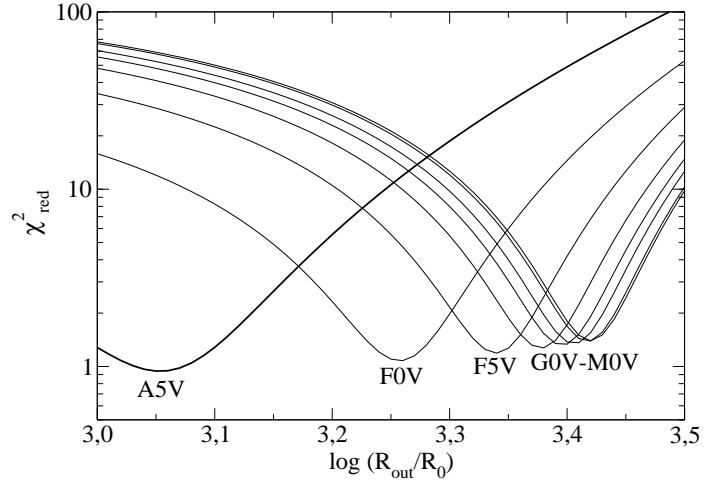}
      \caption{Values of $\chi_{\rm red}^2$ for the range of outer radii and spectral types explored in this paper.
  The SED assuming an A5V companion star provided the smallest value of this statistical indicator.
  The curves corresponding to B and early-type A stars are above and outside the range of this plot.}    
         \label{chi2}
   \end{figure} 
 
    \begin{figure}
   \centering
   \includegraphics[angle=0,width=10.0cm]{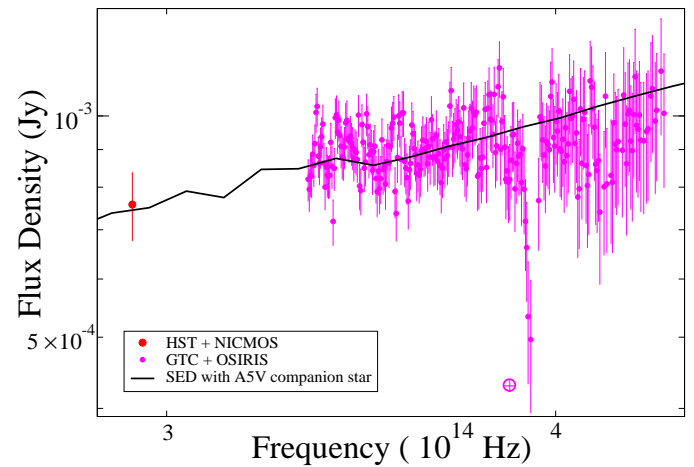}
      \caption{Detailed view of the de-reddened GTC spectrum as a function of frequency and 
      including estimated error bars.
      The solid line is the SED fit corresponding to the $\chi_{\rm red}^2$ minimum in Fig. \ref{chi2} obtained for an A5V star.
      The symbol $\oplus$ marks the strongest telluric absorption that is excluded from the fit.
    }
         \label{sed_zoom}
   \end{figure} 

   To assess the quality of the synthetic SEDs, we computed their respective reduced $\chi_{\rm red}^2$ value excluding the regions of strong telluric absorption.
   The result is presented in Fig. \ref{chi2}, where a minimum value of $\chi_{\rm red}^2 = 0.94$ was achieved in the case of an A5V companion star
   and an outer radius $10^{3.05}$ times the inner value. For a total of 246 degrees of freedom (248 GTC used data points minus two explored parameters),
   the associated non-chance probability is greater than 70\%.
   The corresponding SED fit is represented by the solid line in Fig. \ref{sed_zoom}.  
   The error bars shown take into account both the rms of the GTC spectrum and the 5\% uncertainty in the adopted scaling factor.
   Neither varying the stellar template by a few spectral sub-types nor changing the scaling factor by a few percentage have a strong effect on the best fit result.
   Spectral types later than A5V provide progressively higher $\chi_{\rm red}^2 $ values and lower
   non-chance probabilities  (see Fig. \ref{chi2}  and Table \ref{chi2tab}). Although we cannot fully rule these cooler stars out, 
    the suggestion of an A-type main sequence companion is the best supported by current GTC data.
    
    \begin{table}
\caption{Goodness of the SED fit in the GTC domain.}          
\label{chi2tab}      
\centering                        
\begin{tabular}{c c c c}     
\hline\hline                
Spectral type   &    $\log\left[{R_{\rm out}/R_0}\right]$ &  $\chi_{\rm red}^2$    &  Non-chance   \\
of companion &               &                         &     probability($^{(a)}$) \\
\hline
A5V  &  3.05 &  0.94  &  73.37\%  \\
F0V  &  3.26  &  1.08  &   18.91\%  \\
F5V &   3.34 &    1.19   &  2.25\%  \\
G0V &  3.38 &   1.27   & 0.23\%  \\
G5V &  3.40 &  1.34   &  0.03\%  \\
K0V &   3.41 & 1.36   &  0.01\%  \\
K5V &   3.42 & 1.39   & 0.00\%   \\
M0V &  3.42 & 1.40   &  0.00\%  \\
\hline
\end{tabular}
\tablefoottext{a}{For 246 degrees of freedom.}

\end{table}

    \begin{figure}
   \centering
   \includegraphics[angle=0,width=10.0cm]{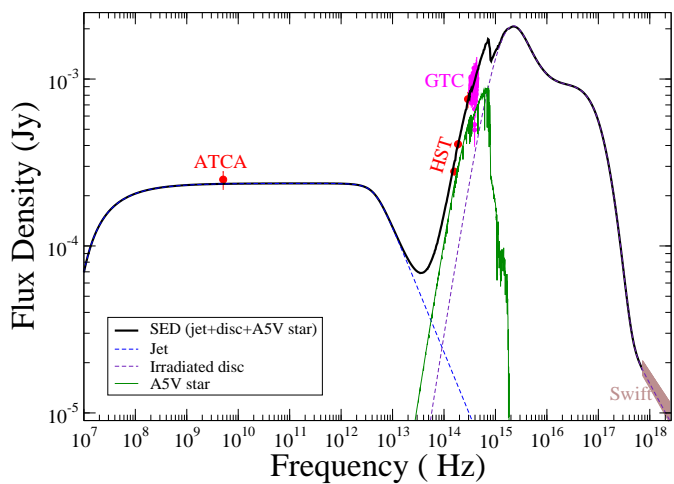}
      \caption{
       Same kind of plot as in Fig. \ref{sed_zoom} but with the SEDs extending from radio to the X-ray domain.
  The solid line is  the same SED best fit using an A5V star in addition to the flat spectrum radio jet and irradiated disc components.
  These components are also plotted separately and identified in the legend.
    }
         \label{sed}
   \end{figure}

    Our tentative interpretation points to the stellar companion and the accretion disc comparably contributing  to the flux in the optical domain,
  where the emission from the synchrotron jet would be negligible. This can be better appreciated in Fig. \ref{sed} that displays a zoomed out version of
  Fig. \ref{sed_zoom} extending from radio to X-ray energies.
 
  It is also important to stress that a mid-A spectral type would render \grs\ a system with possible connections with the sub-class of
  Intermediate Mass X-ray Binaries (IMXBs, see e.g. \citealt{2006csxs.book..623T} for a review). For instance, \grs\ would fit in the spectral type range
  covered by some black hole IMXBs 
  such as 
  \object{GRO J1655$-$40} (spectral type F5IV, \citealt{2002MNRAS.331..351B})
  and
  \object{4U 1543$-$47} (spectral type A2V, \citealt{1998ApJ...499..375O}).
      
\begin{figure}
   \centering
   \includegraphics[angle=0,width=10.0cm]{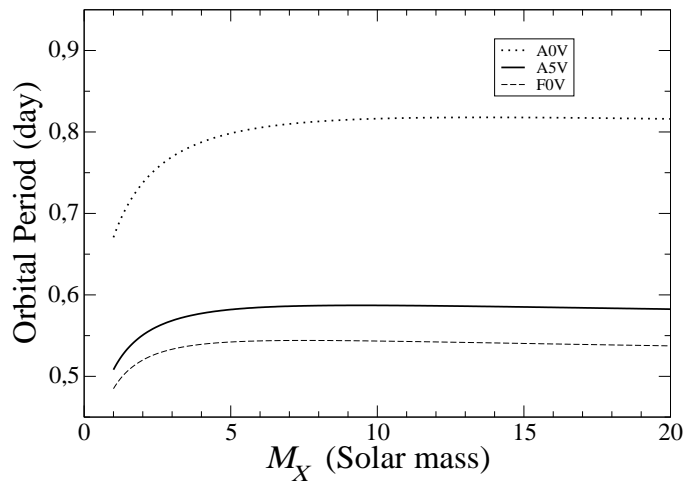}
      \caption{Estimated orbital period of \grs\ as a function of plausible masses $M_X$ of the compact companion and a Roche lobe filling star  
      $\pm5$  spectral sub-types around A5V. The assumed mass and radius of an A5V star
      are $R_* = 2~M_{\odot}$ and $R_* = 1.7~R_{\odot}$ \citep{2000asqu.book.....C}.
       }
       \label{roche}
   \end{figure}

    The suggestion that the stellar companion of \grs\ is a main sequence star instead of a late-type giant implies a re-interpretation of reported X-ray periods of approximately18 d
    previously proposed to represent the orbital cycle \citep{2002ApJ...578L.129S, 2011ApJS..196....6L}, but not yet confirmed.
    Indeed, mass feeding of the compact object by Roche lobe overflow would
    not be feasible for such a long orbital period with a non-giant star. The required orbital period needs to be significantly shorter, as illustrated in Fig. \ref{roche}.
    Here,  the \citet{1983ApJ...268..368E} approximation for the Roche lobe radius  has been used:
   \begin{equation}
   r_L = a \frac{0.49 q^{2/3}}{0.6q^{2/3} +  \log{(1+q^{1/3})}},
   \end{equation}
   where $a$ is the system separation (circular orbit assumed) and $q=M_X/M_*$ is the mass ratio of the primary over the secondary star (the compact object in our case).
    Adopting stellar masses and radii of normal stars from Chapter 15 in  \citet{2000asqu.book.....C},
     together with Kepler's third law, the expected orbital period can be easily computed.
    As a result, an orbital period in the range 0.5-1.0 d is anticipated even accounting for an uncertainty of some spectral sub-types in our SED considerations (see Fig. \ref{roche}).
    In this context, the 18 d period could be perhaps due to a precession motion of the disc. Precession periods in X-ray binaries are usually found to be within
    10-$10^2$ times the orbital value \citep{1998MNRAS.299L..32L}. The ratio 1 to 18 would fit nicely within this range.
  Moreover, a shorter orbital period would also favour a possible disc truncation in agreement with the smaller outer radius 
  provided by our tentative SED fit.

\section{Conclusions}

The optical counterpart of the microquasar \grs\ has been studied with optical spectroscopy using the GTC.
Unfortunately, no emission nor absorption lines were detected.
Despite the challenging difficulties of the observation, the resulting data allow us to put some constraints on the nature of the donor star
in the system.  The shape of the continuum does not agree 
with a dominant contribution of the accretion disc in the optical domain as proposed in previous work.
Instead, the donor star is found to be more 
consistent with a  main sequence object of mid-A spectral type and accounting for a significant fraction of the optical flux.
 If this is correct, \grs\ could be connected with the sub-class of IMXBs.
 Finally, 
 a short orbital period in the 0.5-1.0 d range is predicted in order
to allow Roche lobe overflow. In this context, previously reported X-ray periods of approximately 18 d would correspond to a physical mechanism other than the
orbital cycle such as precession.
Improved observations (e.g.  with adaptive optics or an even larger telescope) will be required to finally reveal photospheric features
in the \grs\ spectrum.

\begin{acknowledgements}
Based on observations made with the Gran Telescopio Canarias (GTC), 
installed in the Spanish Observatorio del Roque de los Muchachos of the Instituto de Astrof\'{\i}sica de Canarias, in the island of La Palma
This work was supported by grant  AYA2013-47447-C3-3-P from the Spanish Ministerio de Econom\'{\i}a y Competitividad (MINECO),
and by the Consejer\'{\i}a de Econom\'{\i}a, Innovaci\'on, Ciencia y Empleo of Junta de Andaluc\'{\i}a
under excellence grant FQM-1343 and
research group FQM-322, as well as FEDER funds.
IRAF is distributed by the National Optical Astronomy Observatory, which is operated by the Association of Universities for Research in Astronomy (AURA) under a cooperative agreement with the National Science Foundation.
\end{acknowledgements}

%
%


\bibliographystyle{aa} 
\bibliography{references.bib} 

\begin{thebibliography}{22}
\expandafter\ifx\csname natexlab\endcsname\relax\def\natexlab#1{#1}\fi

\bibitem[{{Beer} \& {Podsiadlowski}(2002)}]{2002MNRAS.331..351B}
{Beer}, M.~E. \& {Podsiadlowski}, P. 2002, \mnras, 331, 351

\bibitem[{{Bignami} {et~al.}(1991){Bignami}, {Caraveo}, {Mereghetti}, {Paul},
  {Cordier}, {Goldwurm}, {Mandrou}, {Roques}, \&
  {Vedrenne}}]{1991Msngr..66...10B}
{Bignami}, G.~F., {Caraveo}, P.~A., {Mereghetti}, S., {et~al.} 1991, The
  Messenger, 66, 10

\bibitem[{{Bohlin} {et~al.}(1978){Bohlin}, {Savage}, \&
  {Drake}}]{1978ApJ...224..132B}
{Bohlin}, R.~C., {Savage}, B.~D., \& {Drake}, J.~F. 1978, \apj, 224, 132

\bibitem[{{Boyd}(1978)}]{1978JOSA...68..877B}
{Boyd}, R.~W. 1978, Journal of the Optical Society of America (1917-1983), 68,
  877

\bibitem[{{Cox}(2000)}]{2000asqu.book.....C}
{Cox}, A.~N. 2000, {Allen's astrophysical quantities}

\bibitem[{{Eggleton}(1983)}]{1983ApJ...268..368E}
{Eggleton}, P.~P. 1983, \apj, 268, 368

\bibitem[{{Goldwurm} {et~al.}(1994){Goldwurm}, {Cordier}, {Paul}, {Ballet},
  {Bouchet}, {Reques}, {Vedrenne}, {Mandrou}, {Sunyaev}, {Churazov},
  {Gilfanov}, {Finogenov}, {Vikhlinin}, {Dyachkov}, {Khavenson}, \&
  {Kovtunenko}}]{1994Natur.371..589G}
{Goldwurm}, A., {Cordier}, B., {Paul}, J., {et~al.} 1994, \nat, 371, 589

\bibitem[{{Larwood}(1998)}]{1998MNRAS.299L..32L}
{Larwood}, J. 1998, \mnras, 299, L32

\bibitem[{{Levine} {et~al.}(2011){Levine}, {Bradt}, {Chakrabarty}, {Corbet}, \&
  {Harris}}]{2011ApJS..196....6L}
{Levine}, A.~M., {Bradt}, H.~V., {Chakrabarty}, D., {Corbet}, R.~H.~D., \&
  {Harris}, R.~J. 2011, \apjs, 196, 6

\bibitem[{{Luque-Escamilla} {et~al.}(2015){Luque-Escamilla}, {Mart{\'{\i}}}, \&
  {Mart{\'{\i}}nez-Aroza}}]{2015A&A...584A.122L}
{Luque-Escamilla}, P.~L., {Mart{\'{\i}}}, J., \& {Mart{\'{\i}}nez-Aroza}, J.
  2015, \aap, 584, A122

\bibitem[{{Luque-Escamilla} {et~al.}(2014){Luque-Escamilla}, {Mart{\'{\i}}}, \&
  {Mu{\~n}oz-Arjonilla}}]{2014ApJ...797L...1L}
{Luque-Escamilla}, P.~L., {Mart{\'{\i}}}, J., \& {Mu{\~n}oz-Arjonilla},
  {\'A}.~J. 2014, \apjl, 797, L1

\bibitem[{{Mart{\'{\i}}} {et~al.}(2015){Mart{\'{\i}}}, {Luque-Escamilla},
  {Romero}, {S{\'a}nchez-Sutil}, \&
  {Mu{\~n}oz-Arjonilla}}]{2015A&A...578L..11M}
{Mart{\'{\i}}}, J., {Luque-Escamilla}, P.~L., {Romero}, G.~E.,
  {S{\'a}nchez-Sutil}, J.~R., \& {Mu{\~n}oz-Arjonilla}, {\'A}.~J. 2015, \aap,
  578, L11

\bibitem[{{Mart\'{\i}} {et~al.}(1998){Mart\'{\i}}, {Mereghetti}, {Chaty},
  {Mirabel}, {Goldoni}, \& {Rodr\'{\i}guez}}]{1998A&A...338L..95M}
{Mart\'{\i}}, J., {Mereghetti}, S., {Chaty}, S., {et~al.} 1998, \aap, 338, L95

\bibitem[{{Mirabel} {et~al.}(1992){Mirabel}, {Rodr\'{\i}iguez}, {Cordier},
  {Paul}, \& {Lebrun}}]{1992Natur.358..215M}
{Mirabel}, I.~F., {Rodr\'{\i}iguez}, L.~F., {Cordier}, B., {Paul}, J., \&
  {Lebrun}, F. 1992, \nat, 358, 215

\bibitem[{{Mu{\~n}oz-Arjonilla} {et~al.}(2010){Mu{\~n}oz-Arjonilla},
  {Mart{\'{\i}}}, {Luque-Escamilla}, {S{\'a}nchez-Sutil}, {S{\'a}nchez-Ayaso},
  {Combi}, \& {Mirabel}}]{2010A&A...519A..15M}
{Mu{\~n}oz-Arjonilla}, A.~J., {Mart{\'{\i}}}, J., {Luque-Escamilla}, P.~L.,
  {et~al.} 2010, \aap, 519, A15

\bibitem[{{Orosz} {et~al.}(1998){Orosz}, {Jain}, {Bailyn}, {McClintock}, \&
  {Remillard}}]{1998ApJ...499..375O}
{Orosz}, J.~A., {Jain}, R.~K., {Bailyn}, C.~D., {McClintock}, J.~E., \&
  {Remillard}, R.~A. 1998, \apj, 499, 375

\bibitem[{{Rodr\'{\i}guez} {et~al.}(1992){Rodr\'{\i}guez}, {Mirabel}, \&
  {Mart\'{\i}}}]{1992ApJ...401L..15R}
{Rodr\'{\i}guez}, L.~F., {Mirabel}, I.~F., \& {Mart\'{\i}}, J. 1992, \apjl,
  401, L15

\bibitem[{{Rothstein} {et~al.}(2002){Rothstein}, {Eikenberry}, {Chatterjee},
  {Egami}, {Djorgovski}, \& {Heindl}}]{2002ApJ...580L..61R}
{Rothstein}, D.~M., {Eikenberry}, S.~S., {Chatterjee}, S., {et~al.} 2002,
  \apjl, 580, L61

\bibitem[{{Smith} {et~al.}(2002){Smith}, {Heindl}, \&
  {Swank}}]{2002ApJ...578L.129S}
{Smith}, D.~M., {Heindl}, W.~A., \& {Swank}, J.~H. 2002, \apjl, 578, L129

\bibitem[{{Soria} {et~al.}(2011){Soria}, {Broderick}, {Hao}, {Hannikainen},
  {Mehdipour}, {Pottschmidt}, \& {Zhang}}]{2011MNRAS.415..410S}
{Soria}, R., {Broderick}, J.~W., {Hao}, J., {et~al.} 2011, \mnras, 415, 410

\bibitem[{{Sunyaev} {et~al.}(1991){Sunyaev}, {Churazov}, {Gilfanov},
  {Pavlinsky}, {Grebenev}, {Babalyan}, {Dekhanov}, {Yamburenko}, {Bouchet},
  {Niel}, {Roques}, {Mandrou}, {Goldwurm}, {Cordier}, {Laurent}, \&
  {Paul}}]{1991A&A...247L..29S}
{Sunyaev}, R., {Churazov}, E., {Gilfanov}, M., {et~al.} 1991, \aap, 247, L29

\bibitem[{{Tauris} \& {van den Heuvel}(2006)}]{2006csxs.book..623T}
{Tauris}, T.~M. \& {van den Heuvel}, E.~P.~J. 2006, {Formation and evolution of
  compact stellar X-ray sources}, ed. W.~H.~G. {Lewin} \& M.~{van der Klis},
  623--665

\end{thebibliography}


\end{document}